\documentclass[aps,prd,groupedaddress,showpacs,showkeys]{revtex4}
\usepackage{latexsym,epsfig,amsmath,amssymb}
\usepackage{amssymb}
\usepackage{amsmath}
\usepackage{amsfonts}
\usepackage{epsfig}

\newcommand{\seq}{\begin{subequations}}
\newcommand{\sen}{\end{subequations}}
\newcommand{\eq}{\begin{eqnarray}}
\newcommand{\en}{\end{eqnarray}}
\newcommand{\ra}{\rangle}

\def\d{D^0}
\def\db{\bar{D}^{0}}
\def\ds{D^{\ast \, 0}}
\def\dbs{\bar{D}^{\ast \, 0}}

\def\L2{\Lambda^2}

\def\jp{J/\psi}

\begin{document}

\title{Estimate for the $X(3872) \to \gamma \, \jp$ decay width} 

\noindent
\author{Yubing Dong$^{1,2,3}$, 
        Amand  Faessler$^1$,  
        Thomas Gutsche$^1$,
        Valery E. Lyubovitskij$^1$\footnote{On leave of absence
        from Department of Physics, Tomsk State University,
        634050 Tomsk, Russia}
\vspace*{1.2\baselineskip}}

\affiliation{$^1$ Institut f\"ur Theoretische Physik,
Universit\"at T\"ubingen,\\
Auf der Morgenstelle 14, D--72076 T\"ubingen, Germany
\vspace*{1.2\baselineskip} \\
$^2$ Institute of High Energy Physics, Beijing 100049, P. R. China 
\vspace*{1.2\baselineskip} \\ 
$^3$ Theoretical Physics Center for Science Facilities (TPCSF), CAS, 
Beijing 100049, P. R. China\\} 

\date{\today}

\begin{abstract} 

The $X(3872)$ resonance is considered as a hadronic molecule, 
a loosely--bound state of charmed $\d$ and $\ds$ mesons, since 
its mass is very close to the $\ds\db$ threshold. Assuming 
structure and quantum numbers of $X(3872)$ as 
$(\d\dbs - \ds\db)/\sqrt{2}$ and $J^{PC} = 1^{++}$, we calculate 
the $X(3872)\to \gamma \, \jp$ decay width using a phenomenological 
Lagrangian approach. We also estimate the contribution of 
an additional $c\bar c$ component in the $X(3872)$ to this decay
width, which is shown to be suppressed relative to the one of the
molecular configuration.

\end{abstract}



\maketitle

\newpage

\section{Introduction}

During the last years several new meson resonances, whose properties 
cannot be simply explained and understood in conventional quark models, 
have been observed in different experiments. The $X(3872)$ is one of 
such new charmonium states with mass $m_X = 3871.4\pm 0.6$~MeV and 
a narrow width of $\Gamma_X < 2.3$~MeV~\cite{Yao:2006px}. 
The first measurement of X(3872) was carried out by 
the Belle Collaboration 2003~\cite{Choi:2003ue} in $B$--meson decay 
$B^{\pm} \to K^{\pm} X\to K^{\pm} \jp \pi^+\pi^-$. 
Later the existence of the $X(3872)$ was confirmed in the 
experiments of the CDF II~\cite{Acosta:2003zx}, D0~\cite{Abazov:2004kp},  
and BABAR~\cite{Aubert:2004ns} Collaborations. So far, several 
decay modes of the $X(3872)$ into $\pi^+ \, \pi^- \, \jp$, 
$\pi^+ \pi^- \pi^0 \jp$, $\d \db \pi^0$ and $\gamma \jp$ have been 
identified~\cite{Yao:2006px}, which give some constraints on the quantum 
numbers of this state. In particular, the decay mode 
$X(3872) \to \gamma \, \jp$ implies the positive charge parity $C = +$ 
of this resonance. The three--body decays $X(3872)\to\pi^+ \,\pi^-\,\jp$ 
and $X(3872) \to \d \db \pi$ together constrain (or almost fix) 
the spin--parity quantum numbers of $X$ as $J^{PC} = 1^{++}$. 

Several structure interpretations for the $X(3872)$ have been proposed 
in the literature (for a status report see e.g. 
Refs.~\cite{Swanson:2006st,Bauer:2005yu,Voloshin:2007dx}): 
quarkonium ($c \bar c$)~\cite{Barnes:2003vb,Eichten:2004uh,Kong:2006ni}, 
tetraquark (``diquark--antidiquark''~\cite{Maiani:2004vq}--\cite{Chiu:2006hd} 
and ``meson--meson''~\cite{Chiu:2006hd}--\cite{Takeuchi:2007zz} 
configurations), 
hadronic molecule~\cite{Voloshin:1976ap}--\cite{Braaten:2007ft},  
quarkonium--molecule mixtures~\cite{Voloshin:2003nt,Suzuki:2005ha},  
$c \bar c g$ hybrids (gluonic hadrons)~\cite{Giles:1977mp}, 
quarkonium--glueball mixtures~\cite{Seth:2004zb} or even as a dynamical 
``cusp'' related to the near $\d \dbs$ threshold~\cite{Bugg:2004rk}. 
As was already stressed before in the context of molecular 
approaches~\cite{Voloshin:1976ap}--\cite{Braaten:2007ft} the $X(3872)$ 
can be identified with a weakly--bound hadronic molecule whose constituents 
are $D$ and $D^\ast$ mesons. The reason for this natural interpretation is 
that $m_X$ is very close to the $\d \dbs$ threshold and hence is in analogy 
to the deuteron --- a weakly--bound state of proton and neutron. 
Note, that the idea to treat the charmonium states as hadronic molecules 
traces back to Refs.~\cite{Voloshin:1976ap,DeRujula:1976qd}. 
Originally it was proposed that the state $X(3872)$ is a superposition 
of $\d \dbs$ and $\db \ds$ pairs. Later (see e.g. discussion 
in Refs.~\cite{Swanson:2003tb,Voloshin:2004mh,Braaten:2005ai}) also
other structures, such as a charmonium state or even other
meson pair configurations, were discussed in 
addition to the $\d  \dbs +$ charge conjugate (c.c.) 
component.
Note, the possibility that the $X(3872)$ is 
a virtual state is not excluded (see e.g. discussion in 
Ref.~\cite{Braaten:2007ft,Hanhart:2007yq}). In Ref.~\cite{Liu:2008fh} 
(see also~\cite{Liu:2006df,Gamermann:2007fi,Liu:2008du}) 
it was correctly argued that the positive charge parity of the $X(3872)$ 
corresponds to the following wave function: 
$|X(3872)\ra = \frac{1}{\sqrt{2}} ( | \d  \dbs \ra - | \ds \db \ra )$. 
The possibility of two nearly degenerated $X(3872)$ states with positive 
and negative charge parity has been discussed in 
Refs.~\cite{Terasaki:2007uv,Gamermann:2007fi}. 

This paper focuses on the radiative decay $X(3872) \to \gamma \jp$ 
using a phenomenological Lagrangian approach based on the molecular 
$(\d  \dbs - \ds \db)/\sqrt{2}$ structure of the $X(3872)$. The first 
observation of the $X(3872) \to \gamma \jp$ decay mode has been reported 
by the Belle Collaboration~\cite{Abe:2005ix}. In particular, the Belle 
Collaboration indicated the product of branching fractions 
\eq 
{\rm Br}(B \to X K) 
\cdot {\rm Br}(X \to \gamma\jp) 
= (1.8 \pm 0.6 \pm 0.1) \times 10^{-6} \,,  
\en 
and the branching ratio 
\eq\label{data_Belle}
\displaystyle\frac{\Gamma(X\to \gamma\jp)}
{\Gamma(X\to\pi^+\pi^-\jp)}= 0.14 \pm 0.05 \,.
\en 
Later on, the decay mode $X\to \gamma\jp$ was confirmed by 
the BABAR Collaboration~\cite{Aubert:2006aj}. Their result for 
the product of branching fractions was:  
\eq 
{\rm Br}(B^+ \to X K^+) 
\cdot {\rm Br}(X \to \gamma\jp) 
= (3.3 \pm 1.0 \pm 0.3) \times 10^{-6} \,. 
\en 
A theoretical analysis of the $X(3872)\to\gamma\jp$ decay has 
been performed in Refs.~\cite{Barnes:2003vb,Braaten:2003he,%
Swanson:2004pp,Braaten:2005ai}. 
In particular, in Ref.~\cite{Barnes:2003vb} the radiative decays of 
the $X(3872)$ have been considered in detail in the framework of 
a possible $1D$ and $2P$ charmonium interpretation. 
It was found, that the results are very sensitive to the model details
and to the quantum numbers of the $X(3872)$. 
For the assignment $J^{PC} = 0^{++}, 1^{++}$ and $2^{++}$ the following 
results for $\Gamma(X \to \gamma\jp)$ have been obtained: 
1.5 eV, 11 keV and 37.2 keV, respectively. 
In Ref.~\cite{Swanson:2004pp} different radiative decays of the $X(3872)$ 
have been studied using a potential model, where both the charmonium and 
the molecular interpretation of the $X(3872)$ were considered. In the case 
of the charmonium picture the conclusion of Ref.~\cite{Barnes:2003vb} 
related to the strong model dependence of the results was confirmed; 
the use of different potentials and approximations leads to a significant 
variation of the $X \to \gamma\jp$ decay rate.
Accepting the $1^{++}$ quantum numbers of the $X(3872)$ and using 
a potential with Coulomb, linear and smeared hyperfine terms the results 
for $\Gamma(X \to \gamma\jp)$ were given as 139 keV (without the 
zero recoil and dipole approximations) and 71 keV (using the same 
set of approximations as in Ref.~\cite{Barnes:2003vb}). In the case of 
the molecular interpretation two mechanisms, vector meson dominance (VMD) 
(in the $\rho\jp$ and $\omega\jp$ components) and light quark annihilation 
mechanism (in the neutral and charged $D \bar D^\ast$ components), 
have been analyzed. Here the $X \to \gamma\jp$ rate is dominated by 
the VMD mechanism and the 
prediction for the rate $\Gamma(X \to \gamma\jp) = 8$~keV 
is smaller than in the charmonium picture, but by coincidence similar 
to the result of~\cite{Barnes:2003vb}. Therefore, one of the conclusions of 
Ref.~\cite{Swanson:2004pp} was that a more precise measurement of the
$X \to \gamma\jp$ decay properties will shed light on the internal structure 
of the $X(3872)$. In Ref.~\cite{Braaten:2003he} it was argued that the 
radiative decay $X(3872)\to \gamma\jp$ is dominated by the 
$\d \dbs/\db \ds$ components of the $X(3872)$ wave function, when 
the $S$--wave $\d \dbs$ scattering length is very large. 
In Ref.~\cite{Braaten:2005ai} the branching ratio 
${\rm Br}(X\to \gamma\jp)$ has been related to those for 
$X \to \pi^+\pi^- \jp$ and $X \to \pi^+\pi^- \pi^0 \jp$ 
using VMD.  It was concluded that the prediction 
for ${\rm Br}(X\to \gamma\jp)$ is compatible with the Belle 
data~\cite{Abe:2005ix} if the relative phase between the coupling constants 
of $X$ to $J/\psi\,\omega$ and $J/\psi\rho$ pairs is small. 
 
In Refs.~\cite{Faessler:2007gv} 
we developed the formalism for the study of recently observed 
exotic meson states (like $D_{s0}^\ast(2317)$ and $D_{s1}(2460)$) 
as hadronic molecules. In this paper we extend our formalism to 
the decay $X \to \gamma \jp$ assuming that the $X$ is the $S$--wave,  
positive charge parity $(\d  \dbs - \ds \db)/\sqrt{2}$ molecule. 
As for the case of the $D_{s0}^{\ast}$ and  $D_{s1}$ states, 
a composite (molecular) structure of the $X(3872)$ meson is defined 
by the compositeness condition $Z=0$~\cite{Weinberg:1962hj,%
Efimov:1993ei,Anikin:1995cf}  
(see also Refs.~\cite{Faessler:2007gv}). This condition 
implies that the renormalization constant of the hadron wave function 
is set equal to zero or that the hadron exists as a bound state of its 
constituents. The compositeness condition was originally 
applied to the study of the deuteron as a bound state of proton and
neutron~\cite{Weinberg:1962hj}. Then it was extensively used
in low--energy hadron phenomenology as the master equation for the
treatment of mesons and baryons as bound states of light and heavy
constituent quarks (see e.g. Refs.~\cite{Efimov:1993ei,Anikin:1995cf}). 
By constructing a phenomenological Lagrangian including $X$, $\jp$, $\d$ 
and $\ds$ mesonic degrees of freedom and photons we calculate one--loop 
meson diagrams describing the radiative $X \to \gamma \jp$ decay. 
Note, that recently the similar $\gamma \jp$ decay mode of the $X(3700)$,
which is supposed to be a $D \bar D$ bound state, 
has been considered in~\cite{Gamermann:2007bm} using the chiral unitary 
approach (with coupled--channel dynamics). 

In the present manuscript we proceed as follows. First, in Section~II
we discuss the basic notions of our approach. We discuss the effective
mesonic Lagrangian for the treatment of the $X(3872)$ meson 
as a $\d \dbs - \ds \db$ bound state. In addition, we include the
possibility of a $c\bar c$ admixture in the $X(3872)$.
In Section~III we consider the matrix
elements (Feynman diagrams) describing the radiative $\gamma \jp$
decay of a mixed $X(3872)$ configuration, including the molecular and
quarkonia components.
We discuss our numerical results and perform 
a comparison with other theoretical approaches. We show
that the contribution of a possible quarkonium 
component is suppressed relative to the molecular one. 
Finally, in Section~IV we present a short summary of our results.

\section{Approach} 

\subsection{Molecular structure of the $X(3872)$ meson}

In this section we discuss the formalism for the study of the
$X(3872)$ meson interpreted as a hadronic molecule. We consider 
the $X(3872)$ as a $S$--wave molecular state with positive charge 
parity given by the superposition of $\d \dbs$ and $\db \ds$ pairs 
as: 
\eq\label{X_molecule}
|X(3872)\ra = \frac{1}{\sqrt{2}} ( | \d  \dbs \ra - | \ds \db \ra ) \,. 
\en 
We adopt the convention that the spin and parity quantum numbers of
the $X(3872)$ are $J^{PC} = 1^{++}$, 
while its mass we write in the form 
\eq 
m_X = m_{D^0} 
+ m_{D^{\ast 0}} - \epsilon \,, 
\en 
where $m_{D^0} = 1864.85$ MeV and $m_{D^{\ast 0}} = 2006.7$ MeV 
are the $D^0$ and $D^{\ast 0}$ meson masses, respectively; 
$\epsilon > 0$ represents the binding energy. 
Our framework is based on an effective interaction Lagrangian describing 
the couplings of the $X(3872)$ meson to its constituents:
\eq\label{Lagr_X}
{\cal L}_X^M(x) = i \, \frac{g_{_{X}}}{\sqrt{2}} 
\, X^\mu(x) \, \int\! dy \, \Phi_M(y^2) \, 
\biggl( D^0(x+w_{_{D^\ast D}} y) \, \bar D^{\ast \, 0}_\mu(x-w_{_{DD^\ast}} y) 
- \bar D^0(x+w_{_{D^\ast D}} y) \, D^{\ast \, 0}_\mu(x-w_{_{DD^\ast}} y) 
\biggr)\,, 
\en
where the correlation function $\Phi_M$ characterizes the finite size 
of the $X(3872)$ meson as a $(\d \dbs - \ds \db)/\sqrt{2}$ bound state.  
The index $M$ attached to the Lagrangian and the correlation 
function refers to the ``molecular'' configuration. 
In the nonlocal Lagrangian we use the relative Jacobi coordinate $y$  
and  the center--of--mass (CM) coordinate $x$. In Eq.~(\ref{Lagr_X}) 
we introduce the kinematical parameters $w_{ij} = m_i/(m_i + m_j)$. 
A basic requirement for the choice of an explicit form of the correlation
function is that its Fourier transform vanishes sufficiently fast in the 
ultraviolet region of Euclidean space to render the Feynman diagrams 
ultraviolet finite. We adopt the Gaussian form,
$\tilde\Phi_M(p_E^2/\Lambda_M^2) \doteq \exp( - p_E^2/\Lambda_M^2)\,,$
for the Fourier transform of the vertex function, where $p_{E}$ is the
Euclidean Jacobi momentum. Here, $\Lambda_M$
is a size parameter, which characterizes the distribution of 
the $DD^\ast$ constituents inside the molecule.

The coupling constant $g_{_{X}}$ is determined by the compositeness 
condition~\cite{Weinberg:1962hj,Efimov:1993ei,Anikin:1995cf}  
(for an application to $D_{s0}^\ast(2317)$ and $D_{s1}(2460)$ meson 
properties see Ref.~\cite{Faessler:2007gv}.)  
It implies that the renormalization constant of the hadron 
wave function is set equal to zero:
\eq\label{ZX}
Z_X = 1 - (\Sigma^M_X(m_X^2))^\prime = 0 \,.
\en
Here, $(\Sigma^M_X(m_{X}^2))^\prime = g_{_{X}}^2 (\Pi^M_X(m_X^2))^\prime$ 
is the derivative of the transverse part of the mass operator 
$\Sigma^{\mu\nu}_S$, conventionally split into the transverse
$\Sigma_X$ and longitudinal $\Sigma^L_X$  parts as:
\eq
\Sigma^{M, \mu\nu}_X(p) = g^{\mu\nu}_\perp \Sigma_X^M(p^2) 
+ \frac{p^\mu p^\nu}{p^2} \Sigma^{M, L}_X(p^2) \,,
\en
where 
$g^{\mu\nu}_\perp = g^{\mu\nu} - p^\mu p^\nu/p^2$ 
and $g^{\mu\nu}_\perp p_\mu = 0\,.$ 
The mass operator of the $X(3872)$ is described by 
the diagram of Fig.1(a). 

To clarify the physical meaning of the compositeness condition, 
be reminded that the renormalization constant 
$Z_X^{1/2}$ can also be interpreted as the matrix element 
between the physical and the corresponding bare state. 
For the case $Z_X=0$ it follows that the physical state
does not contain the bare one and hence it is exclusively described 
as a bound state of its constituents. As a result of the interaction 
of the $X$ meson with its constituents, the $X$ meson is dressed, 
i.e. its mass and its wave function have to be renormalized. 

Following Eq.~(\ref{ZX}) the coupling constant $g_X$ 
can be expressed in the form: 
\eq\label{gX_coupling}
\frac{1}{g_{_{X}}^2} = \frac{1}{(4 \pi \Lambda_M)^2} \,
\int\limits_0^1 dx \int\limits_0^\infty
\frac{d\alpha \, \alpha \, P(\alpha, x)}{(1 + \alpha)^3}
\,\, \biggl[ \frac{1}{2 \mu_{D^\ast}^2 (1 + \alpha)}
- \frac{d}{dz} \biggr] \tilde \Phi^2_X(z)\,, 
\en
where
\eq 
P(\alpha, x) = \alpha^2 x(1-x) + w_{_{D^\ast \! D}}^2 \alpha x
+ w_{_{DD^\ast}}^2 \alpha (1-x) \,, \ \ \  
z = \mu_{D^\ast}^2 \alpha x + \mu_D^2 \alpha (1-x)
   - \frac{P(\alpha, x)}{1 + \alpha}  \, \mu_X^2 \,, \ \ \  
\mu_i = \frac{m_i}{\Lambda_M}\,. 
\en
Above expressions are valid for any functional form of the correlation
function $\tilde\Phi_M(z)$.

\subsection{$X(3872)$ meson as mixture of molecule and charmonium 
components} 

Following the suggestion (see e.g. discussion 
in Refs.~\cite{Swanson:2003tb,Voloshin:2004mh,Braaten:2005ai}) that 
the $X(3872)$ could be a mixture of molecular and other components -- 
charmonium or even other mesonic pairs, we include the
$c \bar c$ charmonium 
component in the ansatz for the X(3872) structure. Then  
Eq.~(\ref{X_molecule}) is extended as 
\eq\label{X_molecule+cc}
|X(3872)\ra = \frac{\alpha}{\sqrt{2}} ( | \d  \dbs \ra - | \ds \db \ra ) 
+ \beta | c \bar c \ra\,  , 
\en
where the mixing coefficients $\alpha  $ and $\beta $ are kept as
free parameters.
Later on we also present the result for the radiative decay
width of the $X(3872)$ in terms of these free parameters.
The Lagrangian describing the couplings of the $X(3872)$ to its
molecular and charmonium components is written in extension of~(\ref{Lagr_X})
as:  
\eq\label{Lagr_X+cc}
{\cal L}_X(x) &\equiv &{\cal L}_X^{M + c\bar c}(x) 
= g_{_{X}} \, X^\mu(x) \, \biggl( 
\frac{i\alpha}{\sqrt{2}} \, \int\! dy \, \Phi_M(y^2) \, 
\biggl( D^0(x+w_{_{D^\ast D}} y) \, 
\bar D^{\ast \, 0}_\mu(x-w_{_{DD^\ast}} y) \nonumber\\
&-& \bar D^0(x+w_{_{D^\ast D}} y) \, D^{\ast \, 0}_\mu(x-w_{_{DD^\ast}} y) 
\biggr) +\frac{\beta}{m_c} \, \int\! dy \, \Phi_C(y^2) \, 
\bar c(x+y/2) \gamma_\mu \gamma_5 c(x-y/2) \biggr) \, . 
\en
Now the index $C$ indicates quantities related to the charmonium
configuration.
In particular, the correlation function $\Phi_C(y^2)$ characterizes 
the distribution of charm quarks in the $X(3872)$. 
We adopt the Gaussian form for $\Phi_C(y^2)$ function with
$\tilde\Phi_C(p_E^2/\Lambda_C^2) \doteq \exp( - p_E^2/\Lambda_C^2)\,,$ 
where $\Lambda_C$ is a free parameter. 
For dimensional reasons we divide the charmonium component by the
constituent quark mass $m_c$.
We also keep a common coupling constant $g_X$ such that we can
consider the direct limit for the
pure charmonium case: $\alpha \to 0$ and $\beta \to 1$. 

Application of the compositeness condition (now including both components -- 
molecular and charmonium) constrains the parameters $\alpha$ 
and $\beta$ (or their ratio). Now the compositeness condition reads 
\eq 
Z_X = 1 - (\Sigma^M_X(m_X^2))^\prime - (\Sigma^C_X(m_X^2))^\prime = 0 \, ,
\en 
where $(\Sigma^{M,C}_M(m_{X}^2))^\prime$ 
are the derivatives of the transverse part of the
$X(3872)$ mass operator due to the molecular (Fig.1(a)) 
and charmonium (Fig.1(b)) component. 

\subsection{Effective Lagrangian for the radiative decay 
$X \to \gamma \jp$} 

The diagrams contributing to the radiative decay 
$X \to \gamma \jp$ are shown in Fig.2: 
the $\d \ds - \ds$ meson loop diagram [Fig.2(a)] and 
the one involving the $\ds \d - \d$ meson loop [Fig.2(b)] originate
from the molecular $DD^\ast$ component, 
while the quark loop diagram [Fig.2(c)] is related to the contribution 
of the charmonium component. 
The corresponding phenomenological Lagrangian formulated  
in terms of the mesons $X$, $\jp$ (in the Lagrangian we denote 
it by $J_\psi$), $\d$, $\ds$ (for simplicity we suppress the 
charged isopartners), charm quarks and the photon, including free 
and interaction parts, is written as: 
\eq\label{L_full} 
{\cal L}(x) = {\cal L}_{\rm free}(x) + {\cal L}_{\rm int}(x) \,,   
\en 
where 
\seq\label{L_free}
\eq 
{\cal L}_{\rm free}(x) &=&  \sum\limits_{M = X, J_\psi} 
\frac{1}{2} M_\mu (x) ( g^{\mu\nu} [\Box + m_M^2] 
- \partial^\mu \partial^\nu ) M_\nu(x) 
+ \bar c(x) (i \not\!\partial - m_c) c(x) 
- \frac{1}{4} F_{\mu\nu}(x) \, F^{\mu\nu}(x) \nonumber\\[1mm]  
&+&\bar D^{\ast 0}_\mu (x) ( g^{\mu\nu} [\Box + m_{D^{\ast 0}}^2] 
- \partial^\mu \partial^\nu ) D^{\ast 0}_\nu(x) 
- \bar D^0(x) ( \Box + m_{D^0}^2 ) D^0(x) \,,\\[2mm]
{\cal L}_{\rm int}(x) &=& {\cal L}_X(x) + {\cal L}_{J_\psi}(x) + 
{\cal L}_{J_\psi DD}(x) 
+ {\cal L}_{J_\psi D^\ast D^\ast}(x) 
+ {\cal L}_{D^\ast D \, \gamma}(x) 
+ {\cal L}_{cc\gamma}(x)  \,.  
\en 
\sen 
Here, ${\cal L}_{D^\ast D \, \gamma}$ and 
${\cal L}_{cc\gamma}(x)$ are the electromagnetic 
$D^{\ast \, 0}  D^0 \, \gamma$ and $cc\gamma$ interaction Lagrangians:  
\seq 
\eq
{\cal L}_{D^\ast D \, \gamma}(x) &=& \frac{e}{4} \, 
g_{_{D^{\ast \, 0} D^0 \gamma}} \, \epsilon^{\mu\nu\alpha\beta} 
\, F_{\mu\nu}(x) \,  \bar D^{\ast \, 0}_{\alpha\beta}(x) \, D^0(x) 
\, + \, {\rm H.c.}\,,  \label{L_em1}  \\
{\cal L}_{c c \, \gamma}(x) &=& \frac{2 e}{3} A_\mu(x) \bar c(x) 
\gamma^\mu c(x) \,. 
\en 
\sen 
The term ${\cal L}_{J_\psi}(x)$ describes the coupling 
of $J/\Psi$ to its constituent charm quarks: 
\eq 
{\cal L}_{J_\psi}(x) = g_{_{J_\psi}} \, J_\psi^\mu(x) \, 
\bar c(x) \gamma_\mu  c(x) \,, 
\en 
where $g_{_{J_\psi}}$ is the coupling constant. 

${\cal L}_{J_\psi D^0 D^0}$ and 
${\cal L}_{J_\psi D^{\ast \, 0} D^{\ast \, 0}}$ 
are the respective strong interaction Lagrangians
\seq 
\eq
{\cal L}_{J_\psi DD}(x) &=& i g_{_{J_\psi DD}} J_\psi^\mu(x) \, 
\biggl( D^0(x) \partial_\mu \bar D^0(x) - \bar D^0(x) 
\partial_\mu D^0(x) \biggr) \,, \label{L_strong_JDD}\\ 
{\cal L}_{J_\psi D^\ast D^\ast}(x) &=& i g_{_{J_\psi D^\ast D^\ast}} 
\biggl( J_\psi^{\mu\nu}(x) \, \bar D^{\ast \, 0}_\mu \, 
D^{\ast \, 0}_\nu + J_\psi^{\mu}(x) \, \bar D^{\ast \, 0 \, \nu}  \, 
D^{\ast \, 0}_{\mu\nu} 
+ J_\psi^{\nu}(x) \, \bar D^{\ast \, 0}_{\mu\nu} \, 
D^{\ast \, 0 \, \mu} \biggr) \,, \label{L_strong_JDsDs}  
\en 
\sen 
where $F_{\mu\nu} =  \partial_\mu A_\nu - \partial_\nu A_\mu$ 
and $M_{\mu\nu} = \partial_\mu M_\nu - \partial_\nu M_\mu$ 
is the stress tensor of the vector mesons with $M = \ds, J_\psi$. 

The phenomenological strong Lagrangians (\ref{L_strong_JDD}) 
and (\ref{L_strong_JDsDs}), describing 
the couplings of $J/\psi$ to $D(D^\ast)$ mesons, have been intensively 
discussed in the context of $J/\psi$ physics, e.g. charmonium absorption 
by light $\pi$ and $\rho$ mesons, $J/\psi$ production in $D \bar D$ 
interactions (see e.g. Refs.~\cite{Matinyan:1998cb}-\cite{Ivanov:2003ge}) 
and, recently, in the analysis of X(3872) decays using a phenomenological 
meson Lagrangian~\cite{Liu:2006df}. Besides a sign difference 
in the definition of the $g_{_{J_\psi DD}}$ and 
$g_{_{J_\psi D^\ast D^\ast}}$ couplings found in the literature, there is also 
a difference in the structure of the Lagrangian (\ref{L_strong_JDsDs}).  
Here we follow Ref.~\cite{Lin:1999ad} what concerns the explicit form of 
the Lagrangians (\ref{L_strong_JDD}) and (\ref{L_strong_JDsDs}) 
including the sign convention.  

At this level we do not include additional, possible form factors 
at the meson interaction vertices for reasons of simplicity and to have 
less number of free parameters. Such form factors would lead to a further 
reduction of the predicted value for the $X \to \gamma \jp$ decay width. 
The importance of these form factors was mentioned with respect to different 
aspects of charm physics, e.g. to obtain a suppression of the $J/\psi$ 
dissociation cross sections~\cite{Matinyan:1998cb}. This implies that 
our result represents an upper limit for the decay width 
$\Gamma(X \to \gamma \jp)$.  

Values for the coupling constants $g_{_{J_\psi DD}}$ and 
$g_{_{J_\psi D^\ast D^\ast}}$ have been previously deduced using constraints 
of SU(4) flavor, chiral, heavy quark symmetries and in the VMD model~(see 
e.g. discussion in Refs.~\cite{Matinyan:1998cb,Lin:1999ad,Haglin:2000ar}). 
The coupling strengths have also been calculated directly using 
microscopic approaches like QCD sum rules~\cite{Matheus:2002nq}, 
quark models~\cite{Deandrea:2003pv,Ivanov:2003ge}, etc. 
In the present calculation we will use the world averaged values of  
couplings $g_{_{J_\psi DD}}$ and $g_{_{J_\psi D^\ast D^\ast}}$ 
of~\cite{Matinyan:1998cb,Lin:1999ad,Haglin:2000ar,% 
Matheus:2002nq,Deandrea:2003pv,Ivanov:2003ge}:   
\eq 
g_{_{J_\psi DD}} = g_{_{J_\psi D^\ast D^\ast}} = 6.5 \,. 
\en 
Next we comment on the coupling constant $g_{D^{\ast \, 0}D^0\gamma}$, 
where the value is deduced from the data on strong and radiative decays 
of $D^\ast$ mesons. We use the central values for the 
partial decay width $\Gamma(D^{\ast \, +} \to D^0 \pi^+)$ and the  
$D^{\ast \, 0}$ branching ratios of: 
\eq 
\Gamma(D^{\ast \, +} \to D^0 \pi^+) = 65 \text{ keV } \,, \hspace*{.5cm} 
{\rm Br}(D^{\ast \, 0} \to D^0 \pi^0) = 61.9 \% \,, \hspace*{.5cm} 
{\rm Br}(D^{\ast \, 0} \to D^0 \gamma) = 38.1 \% \,. 
\en 
The strong decay width $\Gamma(D^{\ast \, 0} \to D^0 \pi^0)$ is 
deduced by applying  isospin invariance, which relates the 
$D^{\ast \, +}D^0\pi^+$ and $D^{\ast \, 0}D^0\pi^0$ couplings as 
\eq 
\Gamma(D^{\ast \, 0} \to D^0 \pi^0) = \frac{1}{2} \, 
\biggl(\frac{m_{D^{\ast \, +}}}{m_{D^{\ast \, 0}}}\biggr)^5 \, 
\biggr(\frac{\lambda(m_{D^{\ast \, 0}}^2,m_{D^0}^2,m_{\pi^0}^2)} 
{\lambda(m_{D^{\ast \, +}}^2,m_{D^0}^2,m_{\pi^+}^2)}\biggr)^{3/2} \,    
\Gamma(D^{\ast \, +} \to D^0 \pi^+) 
= 42.3 \text{ keV } \,, 
\en 
where $\lambda(x,y,z) = x^2 + y^2 + z^2 - 2 xy - 2 xz - 2 yz$ 
is the K\"allen function.

Then we have the decay width 
$\Gamma(D^{\ast \, 0} \to D^0 \gamma)$ 
which is expressed through the coupling constant 
$g_{D^{\ast \, 0}D^0\gamma}$ as 
\eq\label{gamma_rad} 
\Gamma(D^{\ast \, 0} \to D^0 \gamma) = \frac{\alpha}{24} \, 
g_{_{D^{\ast \, 0}D^0\gamma}}^2 \, m_{D^{\ast \, 0}}^3 \, 
\biggl( 1 - \frac{m_{D^{0}}^2}{m_{D^{\ast \, 0}}^2} \biggr)^3
= 26 \text{ keV } \,. 
\en 
From Eq.~(\ref{gamma_rad}) we finally predict 
\eq 
g_{D^{\ast \, 0}D^0\gamma} \simeq 2 \ \text{ GeV}^{-1} \,. 
\en 
For the mass $m_c$ of the charm quark we choose the value $m_c = m_X/2$. 
The coupling $g_{J_\psi}$ is related to the coupling $f_{J_\psi}$ as 
\eq 
g_{J_\psi} = \frac{2}{3} \frac{m_{J_\psi}}{f_{J_\psi}} \,. 
\en 
The quantity $f_{J_\psi}$ is defined by the decay width 
$J/\Psi \to \gamma \to e^+ e^-$: 
\eq 
\Gamma(J/\Psi \to e^+ e^-) = \frac{16\pi}{27} \, \frac{\alpha^2}{m_{J_\psi}} 
\, f_{J_\psi}^2 \simeq 5.55 \ {\rm keV}\,. 
\en 
Fitting the experimental value with $f_{J_\psi} = 416.5$ MeV
we obtain $g_{J_\psi} \simeq 5$. 
Finally, in our calculation we have the following free parameters: 
the size parameter $\Lambda_M$ in the correlation function $\tilde\Phi_M$, 
describing the distribution of the $DD^\ast$ constituent in the $X(3872)$, 
the size parameter $\Lambda_C$ in the correlation function $\tilde\Phi_C$, 
describing the distribution of the charm quarks in the $X(3872)$ 
and the ratio $R=\beta/\alpha$ of the mixing parameters involving the 
molecular and quarkonia components. 

\section{Radiative decay $X(3872) \to \gamma \jp$} 

\subsection{Matrix element and decay width}

The matrix element describing the radiative $X(3872) \to \gamma \jp$ 
decay is defined in general as follows
\eq\label{Minv_em}
M(X(p) \to \gamma(q) \jp(p^\prime) ) = 
e \, \varepsilon^{m n\rho\sigma} \, \epsilon^\alpha_{_{X}}(p) 
\, \epsilon^\mu_{_{J_\psi}}(p^\prime) \, \epsilon_\rho^\gamma(q) 
\, \frac{q_\sigma}{m_X^2}  \, 
\biggl( A \, g_{\mu n} g_{\alpha m} \, p q + \, B \, g_{\mu n} \, 
p_m q_\alpha \, + \, C \, g_{\alpha m} \, p_n q_\mu\biggr) \; ,
\en
where $A$, $B$ and $C$ are dimensionless couplings, 
$\epsilon^\alpha_{_{X}}$, $\epsilon^\mu_{_{J_\psi}}$ and 
$\epsilon_\rho^\gamma$ 
are the polarization vectors of $X(3872)$, $J/\psi$ and the photon.  

The $X(3872) \to \gamma \jp$ decay width is calculated according to
the expression:
\eq\label{Gamma_X}
\Gamma(X(3872) \to \gamma \jp) \ = \ 
\frac{\alpha}{3} \, \frac{P^{\ast 5}}{m_X^4} \, 
\biggl( ( A + B )^2 + \frac{m_X^2}{m_{J_\psi}^2} ( A + C )^2 
\biggr) \,, 
\en
where $ P^\ast = (m_X^2 -  m_{J_\psi}^2)/(2 m_X)$ is 
the three--momentum of the decay products.  

\subsection{Numerical result and discussion}

First, we discuss our results for the case when the $X(3872)$ 
is a pure molecular state. 
We find that the values of $g_{_{X}}$ are fairly stable 
with respect to a variation of the scale parameter $\Lambda_M$. 
In particular, when varying $\Lambda_M$ from 2 to 3 GeV the coupling 
$g_{_{X}}$ changes from 7.4 to 7.9~GeV. Values for the decay couplings $A^M$, 
$B^M$ and $C^M$ in the same interval of $\Lambda_M$ = 2 -- 3 GeV are: 
\eq 
& &A^M = 2.34 - 3.77 \,, \hspace*{.5cm} 
   B^M = 1.62 - 1.93 \,, \hspace*{.5cm}
   C^M = 3.58 - 4.15 \,, \hspace*{.5cm} {\rm at} \ \epsilon = 0.7 
\ {\rm MeV} 
\,, \nonumber\\
& &A^M = 2.40 - 3.85 \,, \hspace*{.5cm} 
   B^M = 1.65 - 1.96 \,, \hspace*{.5cm}
   C^M = 3.64 - 4.21 \,, \hspace*{.5cm} {\rm at} \ \epsilon = 1 
\hspace*{.4cm} {\rm MeV} \,, \\
& &A^M = 2.49 - 3.97 \,, \hspace*{.5cm} 
   B^M = 1.70 - 1.97 \,, \hspace*{.5cm}
   C^M = 3.74 - 4.30 \,, \hspace*{.5cm} {\rm at} \ \epsilon = 1.5  
\hspace*{.15cm} {\rm MeV} \,, \nonumber 
\en
for various values of the binding energy $\epsilon $. 
Here the superscript $M$ refers to the molecular picture. 
In Table 1, we list our results for the decay width 
$\Gamma(X(3872)\to \gamma\jp)$ at $\epsilon = 0.7, 1, 1.5$ MeV. 
The range of values for our results is due to the variation of 
$\Lambda_M$ from 2 to 3 GeV. Although the resulting decay width is not 
very sensitive to a change in the binding energy $\epsilon$, 
the result depends stronger on the variation of $\Lambda_M$. 
The latter result is consistent with the conclusion of 
Ref.~\cite{Swanson:2004pp}, where the $\Gamma(X(3872) \to \gamma \jp)$ 
decay width is also very sensitive to details of the wave function 
or finite--size effects. We obviously need more data to constrain   
our model parameter $\Lambda_M$. We therefore consider the present 
results as an estimate. For comparison we also present the 
results of Refs.~\cite{Barnes:2003vb,Swanson:2004pp}. 
As was stressed in~\cite{Swanson:2004pp}, in the framework 
of the charmonium picture there is a strong sensitivity to the model 
details, e.g. to the choice of binding potential, leading to a variation 
of the predictions from 11 keV~\cite{Barnes:2003vb} to 
139 keV~\cite{Swanson:2004pp}. On the other hand, our result 
is larger than the prediction $\Gamma(X(3872)\to \gamma\jp) = 8$ keV of 
the other molecular approach~\cite{Swanson:2004pp}. Therefore, a future 
precise measurement of $\Gamma(X(3872) \to \gamma \jp)$ will be a crucial 
check for theoretical approaches.

Next, we consider the admixture of a charmonium component in 
the $X(3872)$. For the following results we fix the binding
energy at $\epsilon = 1$ MeV and use the typical value of 
$\Lambda_C = 2$ GeV. In this case, the coupling constant 
$g_{_{X}}$ is given in terms of the coupling $g_{_{X}}^M$, calculated in 
the ``molecular limit'', by 
\eq 
g_{_{X}} = g_{_{X}}^M 
\ \frac{1}{\alpha^2 + 0.3 \beta^2} \; ,
\en 
where $g_{_{X}}^M = 7.57$ GeV at $\Lambda_M = 2$ GeV and 
$7.63$ GeV at $\Lambda_M = 3$ GeV.
The relative contribution of the molecular and charmonium component
is not sensitive to a variation of the parameter $\Lambda_M$
The limits of a   
pure molecular or charmonium structure are precise with 
$\alpha = 1, \beta = 0$ or $\alpha = 0, \beta = 1$.

For the mixed configuration the results for the decay couplings $A$, 
$B$ and $C$ can be written in terms of the limiting molecular 
case $(A^M, B^M, C^M)$ and the ratio $R = \beta/\alpha$:  
\eq\label{ABC_2GeV} 
A &=& \frac{A^M}{\sqrt{1 + 0.3 R^2}} (1 + 0.364 R )\,,\nonumber\\
B &=& \frac{B^M}{\sqrt{1 + 0.3 R^2}} (1 + 0.014 R )\,,\\
C &=& \frac{B^M}{\sqrt{1 + 0.3 R^2}} (1 - 0.020 R )\, \nonumber
\en 
at $\Lambda_M = 2$ GeV and 
\eq\label{ABC_3GeV}  
A &=& \frac{A^M}{\sqrt{1 + 0.3 R^2}} (1 + 0.228 R )\,,\nonumber\\
B &=& \frac{B^M}{\sqrt{1 + 0.3 R^2}} (1 + 0.012 R )\,,\\
C &=& \frac{B^M}{\sqrt{1 + 0.3 R^2}} (1 - 0.018 R )\, \nonumber 
\en 
for $\Lambda_M = 3$ GeV. 

In the next step we simplify the expression for the $X(3872) \to \gamma \jp$ 
decay width substituting all known parameters and leaving 
the dependence on the couplings $A$, $B$ and $C$: 
\eq\label{Gamma_simp} 
\Gamma(X(3872) \to \gamma \jp) \ = 1.77 \ {\rm keV} \ 
( ( A + B )^2 + 1.562 ( A + C )^2 ) \; . 
\en 
Substituting Eqs.~(\ref{ABC_2GeV}) and (\ref{ABC_3GeV}) into 
the expression (\ref{Gamma_simp})  
we obtain the result for 
$\Gamma(X(3872) \to \gamma \jp)$ in terms of the width
$\Gamma^M(X(3872) \to \gamma \jp)$, 
calculated in the ``molecular limit'', and the ratio $R$ of
mixing parameters:
\eq 
\Gamma(X(3872) \to \gamma \jp) \ = \Gamma^M(X(3872) \to \gamma \jp) \ 
(1 + 0.304 R + 0.025 R^2) 
\en 
at $\Lambda_M = 2$ GeV and 
\eq 
\Gamma(X(3872) \to \gamma \jp) \ = \Gamma^M(X(3872) \to \gamma \jp) \ 
(1 + 0.227 R + 0.013 R^2) 
\en 
at $\Lambda_M = 3$ GeV. Again,
$\Gamma^M(X(3872) \to \gamma \jp) = 129.8$ keV at $\Lambda_M = 2$ GeV 
and 239.1 keV at $\Lambda_M = 3$ GeV (see also Table 1). From the
final expression we conclude
that the contribution of the charmonium component to the 
$X(3872) \to \gamma \jp$ decay width is suppressed relative to the
one of the molecular component. 
 
\section{Summary}

In this paper we have considered the $X(3872)$ resonance with 
$J^{PC} = 1^{++}$ as a hadronic molecule, a loosely--bound state of 
charmed $\d$ and $\ds$ mesons. We also test the possibility 
of the admixture of a charmonium component. Using a phenomenological
Lagrangian approach we have calculated the radiative
$X(3872)\to \gamma \, \jp$ decay width.
We have found that the resulting decay width is not very 
sensitive to a variation of the binding energy $\epsilon$, while it   
depends on the variation of $\Lambda_M$, related to the size of the 
hadronic molecule. We give a final prediction for
$\Gamma(X(3872) \to \gamma \jp)$ in terms of the ratio $R=\beta /\alpha$,
involving the
mixing parameters of the charmonium and molecular components.
We conclude that the contribution of the molecular component dominates
the $X(3872) \to \gamma \jp$ decay width.

\begin{acknowledgments}

This work was supported by the DFG under contracts FA67/31-1 and
GRK683. This work is supported  by the National Sciences Foundations 
No. 10775148 and by CAS grant No. KJCX3-SYW-N2 (YBD). 
This research is also part of the EU Integrated Infrastructure 
Initiative Hadronphysics project under contract number 
RII3-CT-2004-506078 and President grant of Russia
"Scientific Schools"  No. 871.2008.2. 
YBD would like to thank the T\"ubingen theory group for 
its hospitality and Yong-Liang Ma for help.   

\end{acknowledgments}

\begin{table}
\vspace*{2cm}
\begin{center}
{\bf Table 1.}
Decay width of $X(3872) \to \gamma \jp$ in keV.  

\vspace*{.25cm}

\def\arraystretch{1.2}
\begin{tabular}{|l|l|}
\hline
\hspace*{.5cm}
Approach \hspace*{.5cm}
& \hspace*{.5cm}
$\Gamma(X(3872) \to \gamma \jp)$  \hspace*{.5cm}\\
\hline
\,\,\,\,\,
$[c \bar c]\,, \ \ $  Ref.~\cite{Barnes:2003vb} 
& \,\,\,\,\,\,\,\, 11 \,\,\,\,\, \\
\,\,\,\,\,
$[c \bar c]\,, \ \ $ Ref.~\cite{Swanson:2004pp}
\,\,\,\,\, & \,\,\,\,\,\,\,\, 71 \,\,\,\,\, \\
\,\,\,\,\,
$[c \bar c]\,, \ \ $ Ref.~\cite{Swanson:2004pp}
\,\,\,\,\, & \,\,\,\,\,\,\,\, 139 \,\,\,\,\, \\
\,\,\,\,\,
[molecule]$\,, \ \ $ Ref.~\cite{Swanson:2004pp}
\,\,\,\,\, & \,\,\,\,\,\,\,\, 8 \,\,\,\,\, \\
\hline
\,\,\,\,\,    & \,\,\,\,\,\,\,\, 124.8 - 231.3 \ 
($\epsilon = 0.7$ MeV) 
\,\,\,\,\, \\
\,\,\,\,\,
Our results  & \,\,\,\,\,\,\,\, 129.8 - 239.1 \ 
($\epsilon = 1$ MeV) 
\,\,\,\,\, \\
\,\,\,\,\,    & \,\,\,\,\,\,\,\, 138.0 - 251.4 \ 
($\epsilon = 1.5$ MeV) 
\,\,\,\,\, \\
\hline
\end{tabular}
\end{center}
\end{table}

\newpage 

\begin{figure}

\vspace*{2cm} 

\begin{center}
\epsfig{file=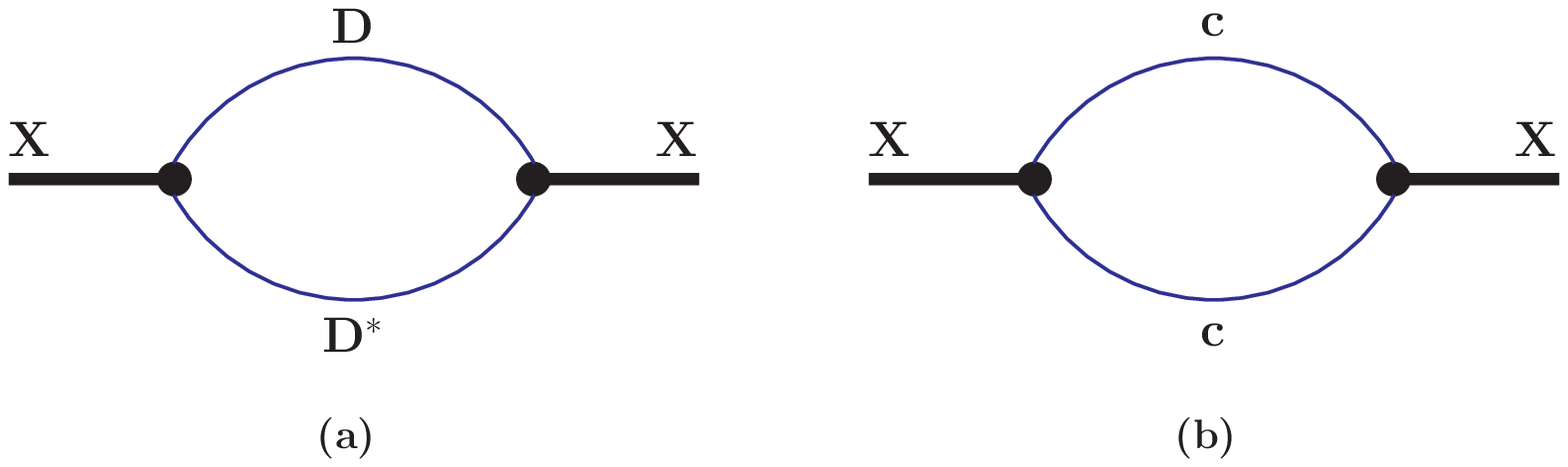, scale=.75}
\end{center}
\caption{Diagrams contributing to the mass operator of the $X(3872)$ meson.}

\vspace*{2cm}

\begin{center}
\epsfig{file=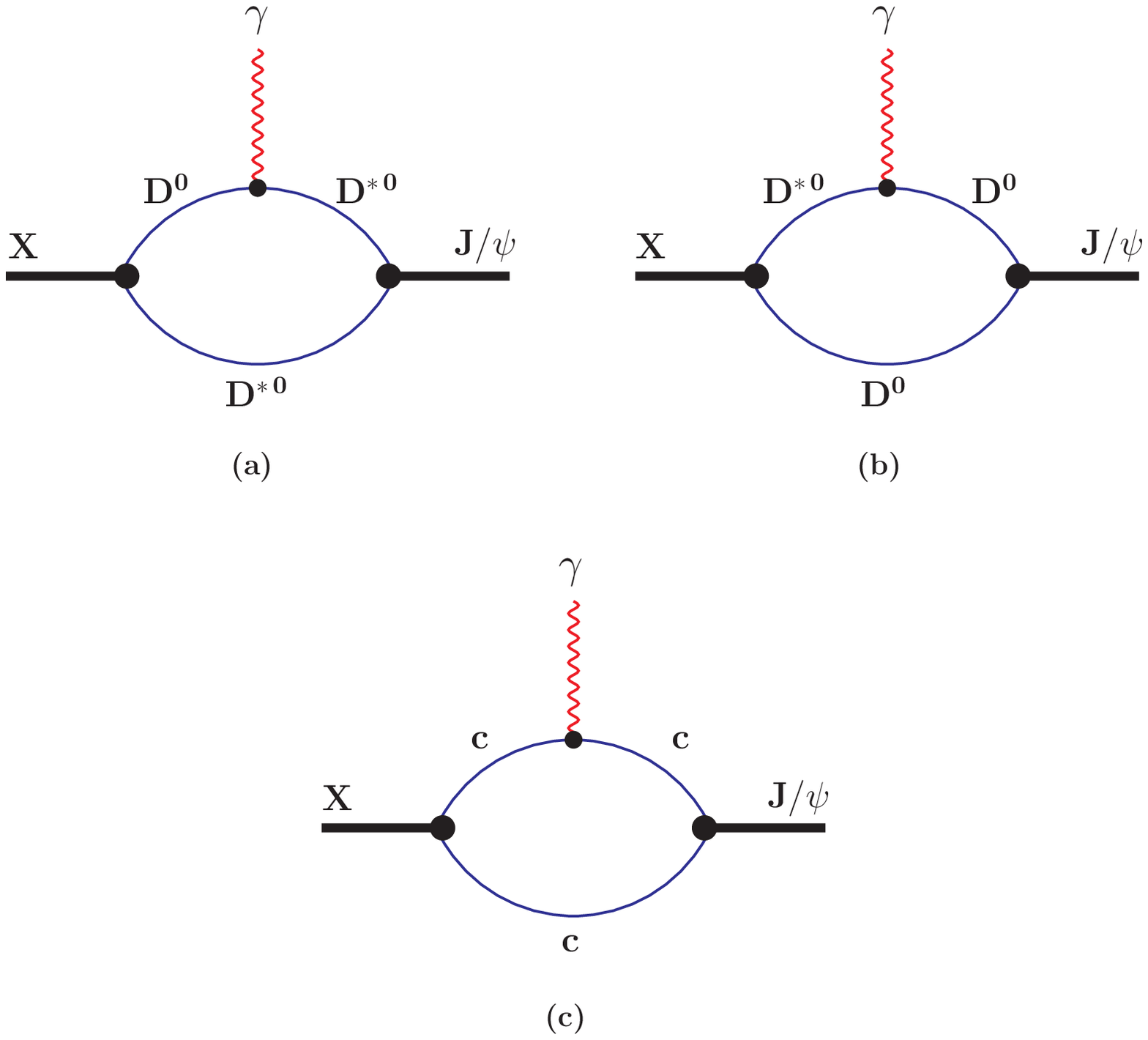, scale=.75}
\end{center}
\caption{Diagrams contributing to the radiative transition
$X(3872) \to \gamma \jp$.}
\end{figure}

\end{document}